\documentclass[3p,times,procedia]{elsarticle}
\usepackage{nupha_ecrc}

\volume{00}

\firstpage{1}

\journalname{Nuclear Physics A}

\runauth{}

\jid{nupha}

\jnltitlelogo{Nuclear Physics A}

\usepackage{amssymb}
\usepackage[figuresright]{rotating}

\newcommand{\pp}{p+p}
\newcommand{\pA}{p+A}
\newcommand{\AuAu}{Au+Au}
\newcommand{\PbPb}{Pb+Pb}
\newcommand{\sqrts}{\ensuremath{\sqrt{s}}}
\newcommand{\sqrtsNN}{\ensuremath{\sqrt{s_{\rm{NN}}}}}
\newcommand{\gev}{GeV/$c$}

\newcommand{\jpsi}{\ensuremath{J/\psi}}
\newcommand{\psiprime}{\ensuremath{\psi}(2S)}
\newcommand{\ups}{\ensuremath{\Upsilon}}
\newcommand{\pT}{\ensuremath{p_{\rm{T}}}}
\newcommand{\RAA}{\ensuremath{R_{\rm{AA}}}}
\newcommand{\RpA}{\ensuremath{R_{\rm{pA}}}}	

\newcommand{\RpPb}{\ensuremath{R_{\rm{pPb}}}}
\newcommand{\vtwo}{\ensuremath{v_{2}}}

\newcommand{\npart}{\ensuremath{N_{\rm{part}}}}

\begin{document}

\begin{frontmatter}



%% Instructions from Editor: Please use the following \dochead only in the preprint version (e-print arXiv etc.); 
%% use empty \dochead{} when submitting to Nuclear Physics A!
\dochead{XXVIIth International Conference on Ultrarelativistic Nucleus-Nucleus Collisions\\ (Quark Matter 2018)}
%% e.g. \dochead{17th International Conference on Dynamical Processes in Excited States of Solids}

\title{Quarkonium production in nuclear collisions}


\author{Rongrong Ma}

\address{Brookhaven National Laboratory, Upton, NY 11973, USA}

\begin{abstract}
In these proceedings, an overview of recent quarkonium measurements in nuclear collisions carried out at both RHIC and LHC is presented. In \pp\ collisions, despite theoretical progresses made in understanding the production mechanisms for quarkonia, a complete picture is still yet to be achieved. In \pA\ collisions where measurements are done to quantify the cold nuclear matter effects, significant suppression of quarkonium production is observed for low transverse momentum region at both forward- and mid-rapidities. Furthermore, results from A+A collisions show the interplay of dissociation and regeneration contributions for charmonium at different collision energies, while a correspondence between the suppression level and the binding energy is observed for bottomonium family. Comparisons to experimental data provide stringent tests to model calculations, and help constrain the temperature of the deconfined medium created in heavy-ion collisions.
\end{abstract}

\begin{keyword}
Quark-gluon plasma \sep deconfinement \sep quarkonium \sep dissociation \sep sequential suppression


\end{keyword}

\end{frontmatter}

% \linenumbers

\section{Introduction}
\label{sec:Intro}
One of the main goals of high-energy nuclear physics is to create the so-call Quark-Gluon Plasma (QGP) and understand its properties utilizing ultra-relativistic heavy-ion collisions. The QGP consists of deconfined quarks and gluons which are freed up from hadrons during the phase transition when the system exceeds the critical temperature. Since this new state of matter has an extremely short lifetime ($\sim\mathcal{O}(1)$ fm/c), internal probes produced along with the medium are exclusively used to study its properties. 

Among various probes used, quarkonia have played a unique role since they usually enter the QGP as a bound state instead of deconfined partons. Furthermore, the heavy quark pairs constituting quarkonia are predominantly produced at early stages of heavy-ion collisions due to their large masses, and therefore they imprint the information throughout the entire evolution of the medium. As suggested more than 30 years ago, the potential that binds the heavy quark and anti-quark pair within a quarkonium can get color-screened in the deconfined medium, leading to a dissociation of the quarkonium state once the Debye radius that is inversely proportional to the medium temperature becomes smaller than the quarkonium radius. Therefore, the \jpsi\ suppression in heavy-ion collisions was proposed as a direct evidence of the QGP formation \cite{Matsui:1986dk}. As the quarkonium dissociation is sensitive to the medium temperature, different quarkonium states of different binding energies are expected to dissociate at different temperatures. Measurement of this sequential suppression can help constrain the medium temperature.

\section{Quarkonium measurements}
\subsection{p+p collisions}
The puzzle of the quarkonium production mechanism in elementary collisions is still not fully solved despite many years of research \cite{Andronic:2015wma}. Continuous efforts are being made from both experimental and theoretical communities in order to achieve a complete understanding of the quarkonium production. 

\begin{figure}[htbp]
\begin{minipage}{0.49\linewidth}
\centerline{\includegraphics[width=0.8\linewidth]{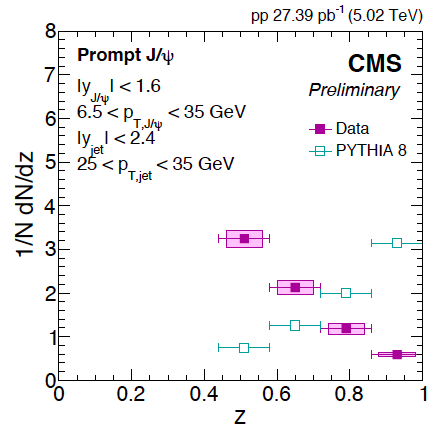}}
\end{minipage}
\begin{minipage}{0.49\linewidth}
\centerline{\includegraphics[width=0.95\linewidth]{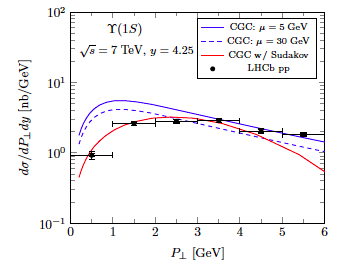}}
\end{minipage}
\caption[]{Left: distribution of prompt \jpsi\ fragmentation function within reconstructed jets of radius $R = 0.4$ for \pp\ collisions at \sqrts\ = 5.02 TeV. Data are shown as filled squares, while the PYTHIA calculation is shown in open squares. Right: calculation of inclusive \ups(1S) cross-section as a function of \pT\ in \sqrts\ = 7 TeV \pp\ collisions using the CGC formalism with (blue) and without (red) Sudakov resummation. Experimental results are shown as filled circles \cite{LHCb:2012aa}. }
\label{fig:JpsiInpp}
\end{figure}

The left panel of Fig. \ref{fig:JpsiInpp} shows the latest measurement of the prompt \jpsi\ fragmentation function within a jet in \pp\ collisions at \sqrts\ = 5.02 TeV by the CMS experiment. 
%-----------------------------
%-----------------------------
The transverse momentum (\pT) of the \jpsi\ meson is above 6.5 \gev, while the jets are reconstructed using the anti-$k_{\rm{T}}$ algorithm \cite{Cacciari:2008gp} with the resolution parameter of $R=0.4$ and are restricted to be within $25<\pT^{\rm{jet}}<35$ \gev. The experimental results are displayed as filled squares, which increase rapidly towards small $z=\pT^{\jpsi}/\pT^{\rm{jet}}$ indicating substantial hadronic activities around \jpsi\ during production. On the other hand, a PYTHIA calculation (open squares) shows a peak around $z=1$, which means that \jpsi's are more likely to be produced without any company of other hadrons. The drastic difference between data and PYTHIA calculation imposes a great challenge as well as an opportunity for the understanding of the \jpsi\ production mechanism. 

Figure 1, right panel, shows the inclusive \ups(1S) production cross-section as a function of \pT\ measured in \sqrts\ = 7 TeV \pp\ collisions by the LHCb collaboration \cite{LHCb:2012aa}. A theoretical calculation employing the Color Glass Condensate (CGC) effective theory to cope with the gluon saturation at small $x$ is shown for comparison as the blue curves \cite{Watanabe:2015yca}. It significantly overshoots the cross-section at low \ups(1S) \pT, where the phase space for gluon showing is greatly enhanced. This is cured by introducing the Sudakov resummation on top of the small-$x$ resummation \cite{Watanabe:2015yca}. The result is shown as the red curve in the panel, and agrees with data quite well. 

\subsection{p+A collisions}
A prevailing approach to study of the QGP properties is to compare similar measurements done in A+A and p+p collisions. However, due to the presence of nuclei in the collision, other effects not related to the creation of the hot medium might also play a role. These effects are usually referred as the Cold Nuclear Matter (CNM) effects, and quantified by carrying out similar measurements in p+A collisions, where the QGP is not expected to be produced. 

\begin{figure}[htbp]
\begin{minipage}{0.49\linewidth}
\centerline{\includegraphics[width=0.95\linewidth]{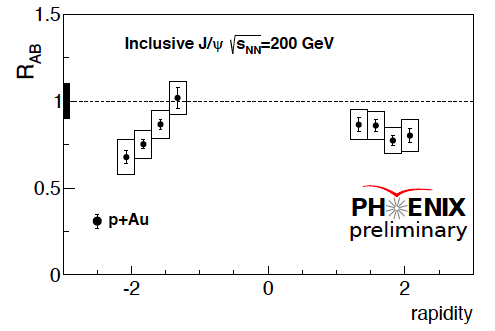}}
\end{minipage}
\begin{minipage}{0.49\linewidth}
\centerline{\includegraphics[width=0.9\linewidth]{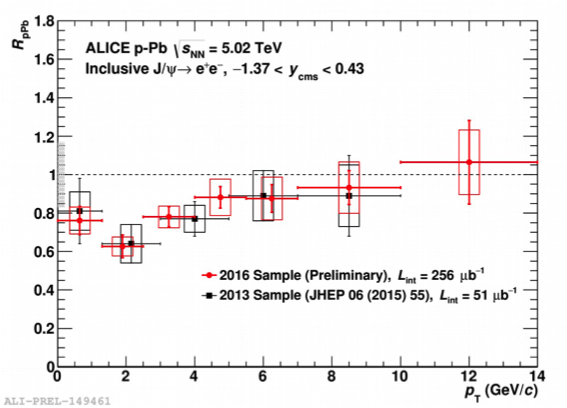}}
\end{minipage}
\caption[]{Measurements of \jpsi\ \RpA\ as a function of rapidity in 200 GeV p+Au collisions (left) and as a function of \pT\ in \sqrtsNN\ = 5.02 TeV (right) p+Pb collisions.}
\label{fig:JpsiRpa}
\end{figure}

\begin{figure}[htbp]
\begin{minipage}{0.4\linewidth}
\centerline{\includegraphics[width=0.95\linewidth]{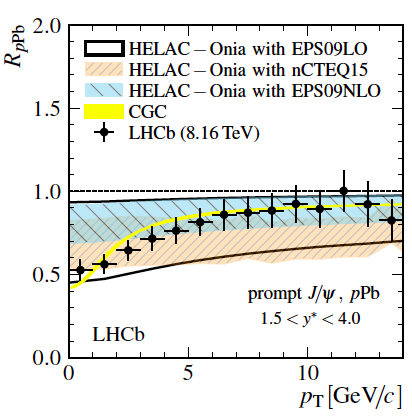}}
\end{minipage}
\begin{minipage}{0.5\linewidth}
\centerline{\includegraphics[width=0.95\linewidth]{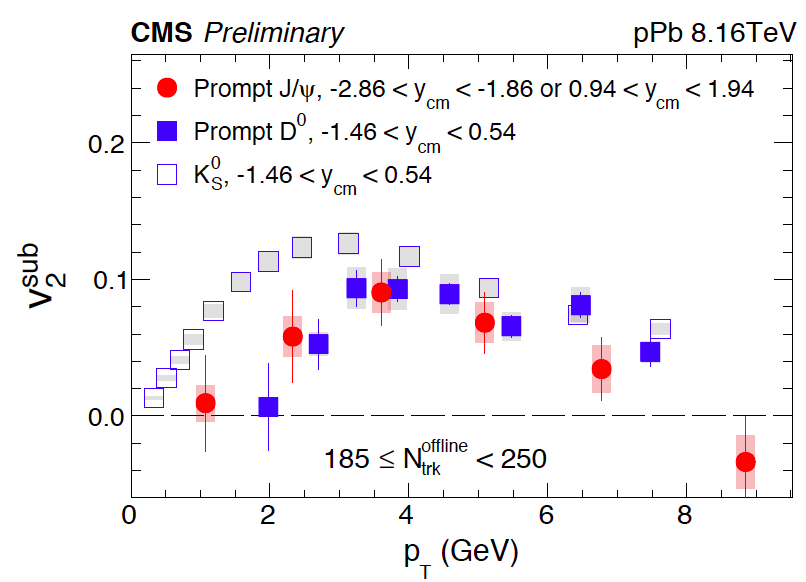}}
\end{minipage}
\caption[]{Prompt \jpsi\ \RpA\ (left) \cite{Aaij:2017cqq} and \vtwo\ (right) as a function of \pT\ in p+Pb collisions at \sqrtsNN\ = 8.16 TeV.}
\label{fig:JpsiV2InpA}
\end{figure}

Figure \ref{fig:JpsiRpa} and the left panel of Fig. \ref{fig:JpsiV2InpA} show the nuclear modification factor (\RpA) for \jpsi\ mesons as a function of rapidity ($y$) in \sqrtsNN\ = 200 GeV p+Au collisions, as a function of \pT\ in \sqrtsNN\ = 5.02 TeV and 8.16 TeV \cite{Aaij:2017cqq} p+Pb collisions, respectively.
%-----------------------------
%-----------------------------
In p+Au collisions, a suppression of the \jpsi\ yield is seen at forward rapidity, i.e. p-going direction, which is consistent with the expectation of the shadowing effect in the nuclear parton distribution function (nPDF). At backward rapidity, the \jpsi\ yield also seems to be suppressed, the cause of which still remains to be understood. On the other hand, up to 50\% suppression is present for low-\pT\ \jpsi\ at mid- and forward-rapidity in p+Pb collisions at both \sqrtsNN\ = 5.02 TeV and 8.16 TeV. The suppression gradually goes away as \jpsi\ \pT\ increases. At \sqrtsNN\ = 8.16 TeV, model calculations incorporating different nPDF sets as well as using the CGC effective theory are consistent with data. In particular, the total experimental uncertainties are smaller than those from the nPDF sets, which opens the door to constraining the gluon distribution in a nucleus using these data. A complementary measurement of prompt \jpsi\ \vtwo\ in high-multiplicity p+Pb collisions at  \sqrtsNN\ = 8.16 TeV is shown in the right panel of Fig. \ref{fig:JpsiV2InpA} (red circles), compared to similar results for $D^{0}$ and $K_{S}^{0}$ mesons.
%-----------------------------
%-----------------------------
The \jpsi\ \vtwo\ is significantly above zero within $2<\pT^{\jpsi}<6$ \gev, and the magnitude is very similar to that observed for $D^{0}$ despite the slightly different rapidity coverages for the two measurements. The results presented in Figs. \ref{fig:JpsiRpa} and \ref{fig:JpsiV2InpA} follow the observations for other hard probes in p+Pb collisions, i.e. they all show significant \vtwo\ but little yield suppression. Given that a \jpsi\ meson is color neutral and does not interact strongly, this behavior might arise from certain initial-state effects that affects all the hard probes in a similar fashion. 

\begin{figure}[htbp]
\begin{minipage}{0.49\linewidth}
\centerline{\includegraphics[width=0.95\linewidth]{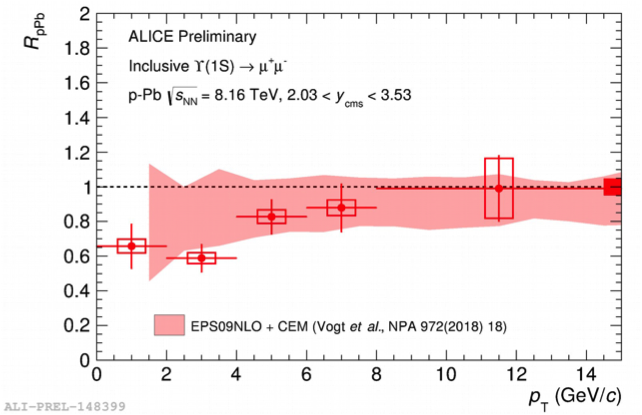}}
\end{minipage}
\begin{minipage}{0.49\linewidth}
\centerline{\includegraphics[width=0.95\linewidth]{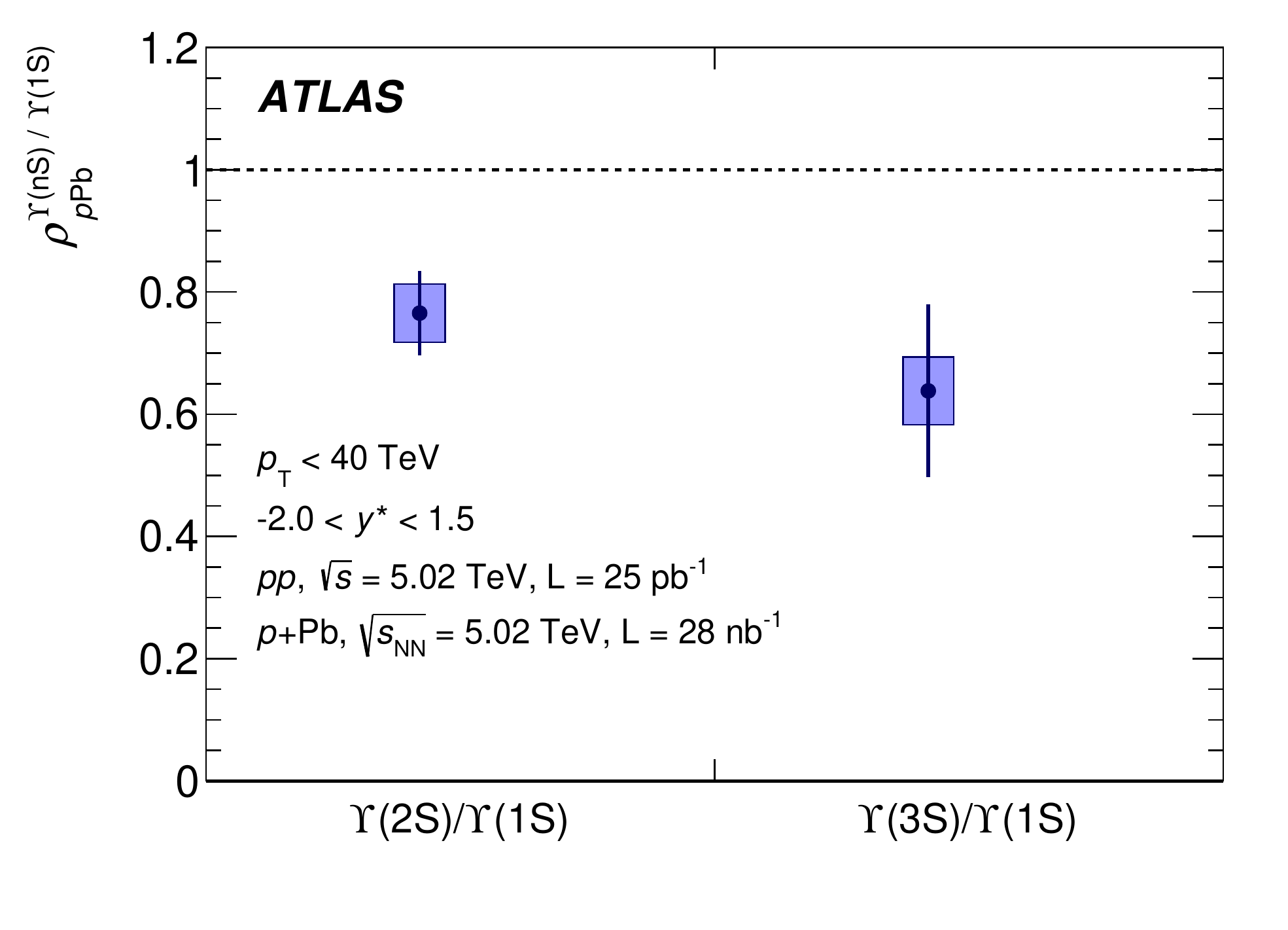}}
\end{minipage}
\caption[]{Left: inclusive \ups(1S) \RpPb\ as a function of \pT\ at forward rapidity in p+Pb collisions at \sqrtsNN\ = 8.16 TeV. A theoretical calculation incorporating the EPS09NLO nPDF is shown as the red band for comparison. Right: double ratios of the \RpPb\ for excited \ups\ states to that of the ground state in \sqrtsNN\ = 5.02 TeV p+Pb collisions \cite{Aaboud:2017cif}. }
\label{fig:UpsRpa}
\end{figure}

The bottomonium family is also suppressed in high-energy p+Pb collisions at the LHC as shown in Fig. \ref{fig:UpsRpa}.
%-----------------------------
%-----------------------------
The data points in the left panel are for inclusive \ups(1S) \RpA\ at the forward rapidity measured by the ALICE experiment, where up to about 40\% suppression is seen at low \pT. A Color Evaporation Model (CEM) calculation incorporating the EPS09NLO nPDF set is consistent with the data within large uncertainties. Double ratios of the \RpA\ for both \ups(2S) and \ups(3S) to that of \ups(1S) are shown in the right panel. Additional suppression is seen for excited \ups\ states, which is usually attributed to final-state effects, such as the co-mover interaction \cite{Ferreiro:2012rq}, since the initial-state effects are the same for different \ups\ states.

\subsection{A+A collisions}
The usage of quarkonia to probe the properties of the QGP is centered on the measurements of the sequential suppression pattern for different quarkonium states with different binding energies, from which one hopes to constrain the medium temperature. 

\subsubsection{Charmonium}
\begin{figure}[htbp]
\begin{minipage}{0.49\linewidth}
\centerline{\includegraphics[width=0.95\linewidth]{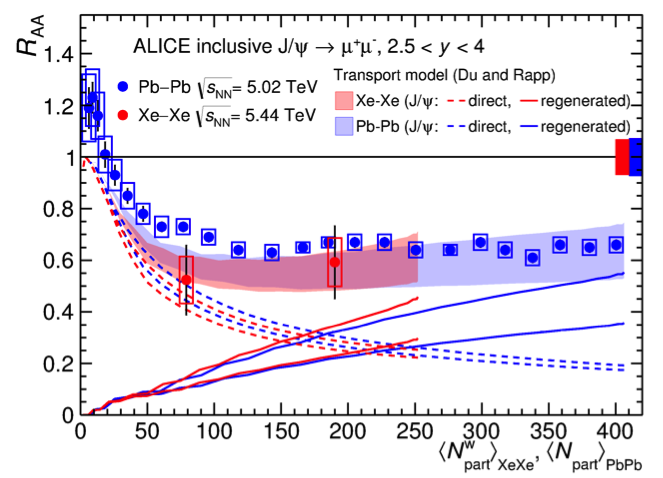}}
\end{minipage}
\begin{minipage}{0.49\linewidth}
\centerline{\includegraphics[width=0.75\linewidth]{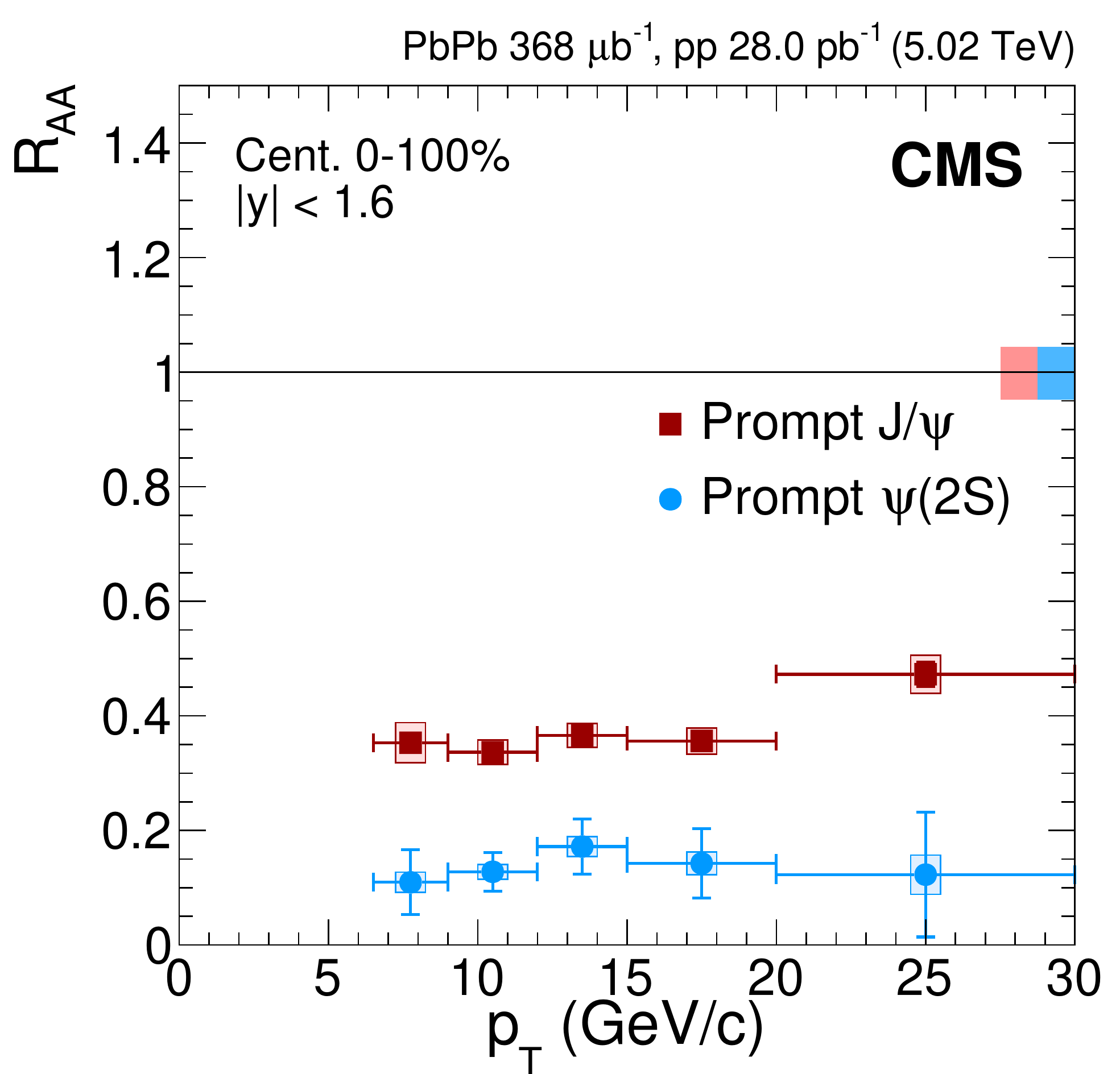}}
\end{minipage}
\caption[]{Left: measurements of \jpsi\ \RAA\ as a function \npart\ in Pb+Pb collisions at \sqrtsNN\ = 5.02 TeV and Xe+Xe collisions at \sqrtsNN\ = 5.44 TeV \cite{Adam:2016rdg, Acharya:2018jvc}. Transport model calculations, including both direct and regenerated \jpsi, are shown for comparison \cite{Zhao:2011cv, Du:2015wha}. Right: comparison of \RAA\ for prompt \jpsi\ and \psiprime\ as a function of \pT\ in minimum-bias \PbPb\ collisions at \sqrtsNN\ = 5.02 TeV \cite{Sirunyan:2017isk}. }
\label{fig:JpsiRaa}
\end{figure}

The new measurement of the inclusive \jpsi\ \RAA\ as a function of centrality in Xe+Xe collisions at \sqrtsNN\ = 5.44 TeV \cite{Acharya:2018jvc} is shown in the left panel of Fig. \ref{fig:JpsiRaa} as red points. The level of suppression is compatible to that measured in \sqrtsNN\ = 5.02 TeV Pb+Pb collisions \cite{Adam:2016rdg}. 
%-----------------------------
%-----------------------------
Transport model calculations \cite{Zhao:2011cv, Du:2015wha}, including both direct and regenerated \jpsi\  (dashed and solid lines in the figure), can qualitatively describe the centrality dependence in both collision systems. This confirms the importance of regeneration as an additional production mechanism in heavy-ion collisions when the total $c\bar{c}$ cross-section becomes substantial. The right panel of Fig. \ref{fig:JpsiRaa} shows the prompt \jpsi\ and \psiprime\ \RAA\ as a function of \pT\ above 6.5 \gev. Since both the CNM effects and the regeneration contribution are expected to be minimal at high \pT, $\RAA\approx0.35$ for \jpsi\ is a clear evidence of the QGP formation since the suppression results mainly from the quarkonium dissociation in the medium due to the color screening effect. Furthermore, as an excited state, the \psiprime\ meson is much more suppressed than \jpsi, consistent with the expectation of sequential suppression as \psiprime\ has a much smaller binding energy. 

\begin{figure}[htbp]
\begin{minipage}{0.49\linewidth}
\centerline{\includegraphics[width=0.75\linewidth]{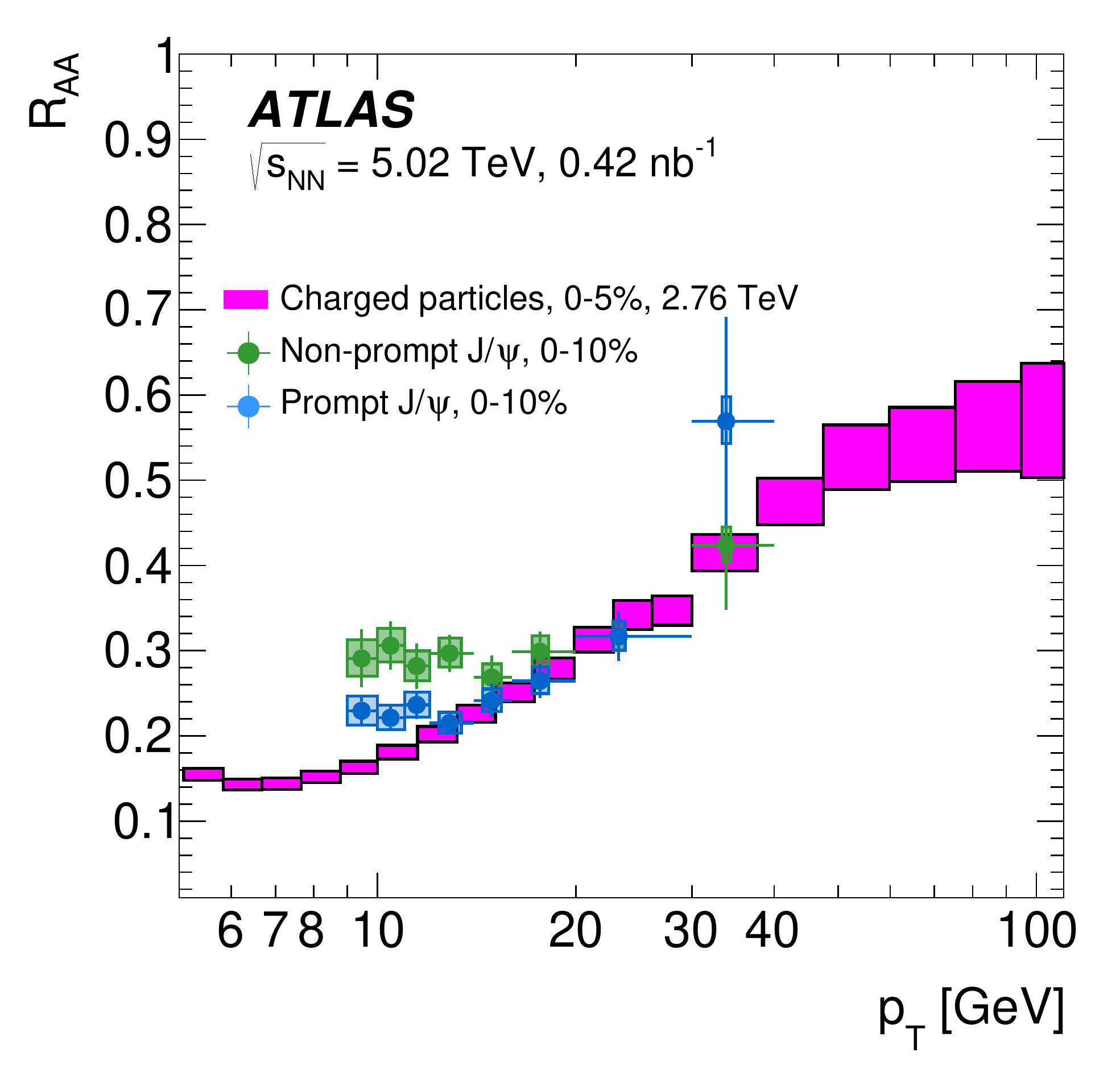}}
\end{minipage}
\begin{minipage}{0.49\linewidth}
\centerline{\includegraphics[width=0.95\linewidth]{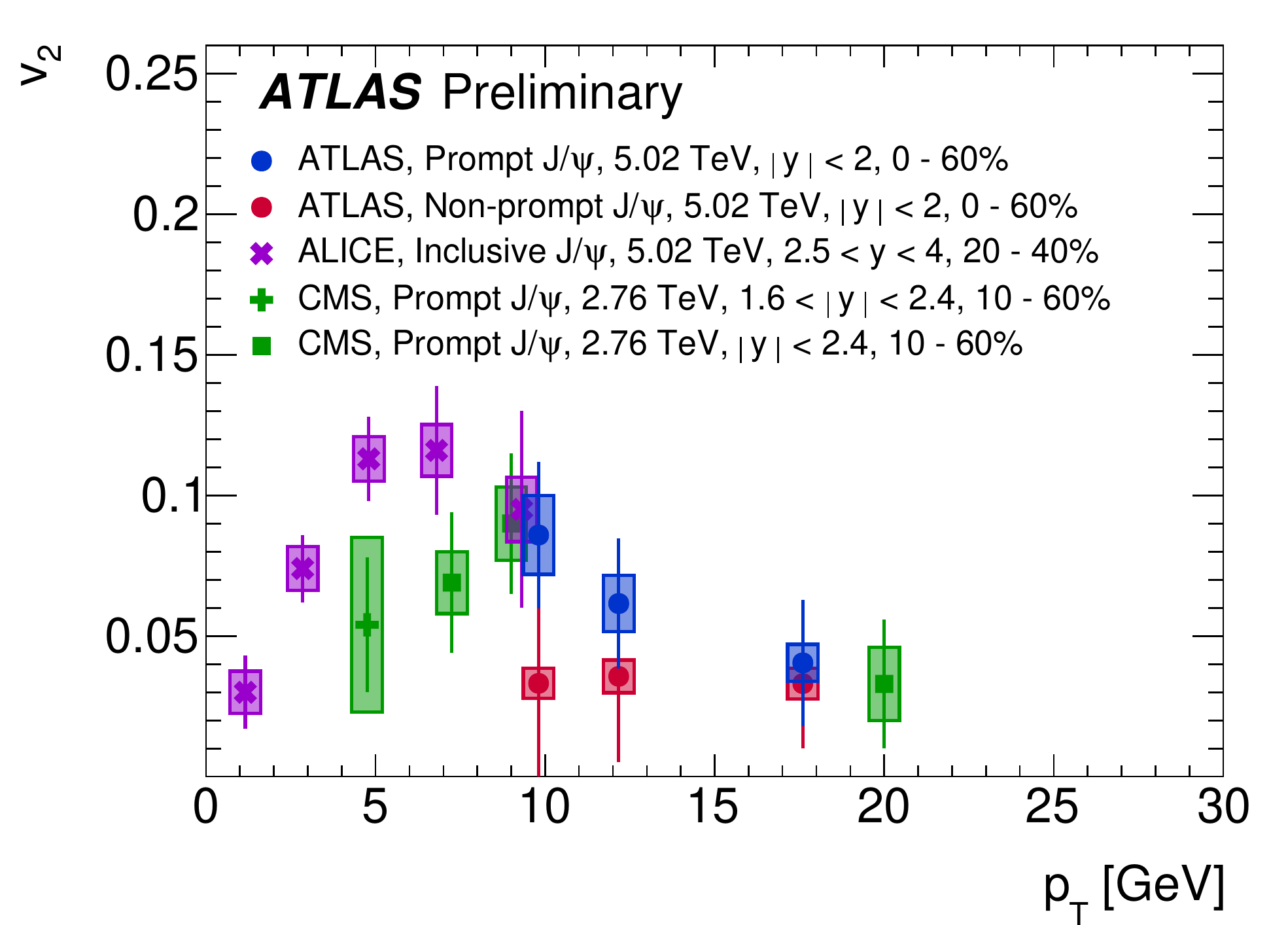}}
\end{minipage}
\caption[]{Left: prompt \jpsi\ \RAA\ above 9 \gev\ compared to those of non-prompt \jpsi\ and charged hadrons in \sqrtsNN\ = 5.02 TeV \PbPb\ collisions \cite{Aaboud:2018quy}. Right: compilation of \jpsi\ \vtwo\ vs. \pT\ measured by multiple experiments at the LHC in \PbPb\ collisions \cite{ATLAS:2018jkw}. }
\label{fig:JpsiV2AA}
\end{figure}

The prompt \jpsi\ \RAA\ measured at even higher \pT\ (up to 40 \gev) is shown in the left panel of Fig. \ref{fig:JpsiV2AA} for 0-10\% most central Pb+Pb collisions at \sqrtsNN\ = 5.02 TeV. Similar results for charged particles as well as for non-prompt \jpsi\ from B-hadron decays are also shown for comparison. Surprisingly, the \jpsi\ \RAA\ above 10 \gev\ or so tracks the charged hadron \RAA\ closely, indicating that the parton energy loss, which is responsible for the charged hadron suppression at high \pT, might also play an important role for very high-\pT\ \jpsi\ suppression if they are mainly produced through parton fragmentation outside of the medium. As an independent handle, the \jpsi\ \vtwo\ in \PbPb\ collisions measured by several experiments at the LHC is shown in the right panel of Fig. \ref{fig:JpsiV2AA}. The result covers $0<\pT^{\jpsi}< 20$ \gev, and a non-zero \vtwo\ persists up to the highest \pT\ region. While the large \jpsi\ \vtwo\ at low to intermediate \pT\ region can be explained by a large contribution of regenerated \jpsi\ which inherits the flow of the constituent charm quarks thermalized in the medium, the non-zero \vtwo\ at very high \pT\ is likely to come from other mechanisms, for example the path-length dependence of the parton energy loss. Even though the parton energy loss is a plausible explanation for the \jpsi\ \RAA\ and \vtwo\ results at very high \pT, its applicability needs to be further scrutinized such that a consistent picture could emerge for the behavior of all the charmonium states. 
%-----------------------------
%-----------------------------

\subsubsection{Bottomonium}
The three \ups\ states, whose binding energies differ by more than a factor of 5, make up a perfect test ground for the relationship between the yield suppression and the medium temperature. 

\begin{figure}[htbp]
\begin{minipage}{0.33\linewidth}
\centerline{\includegraphics[width=0.95\linewidth]{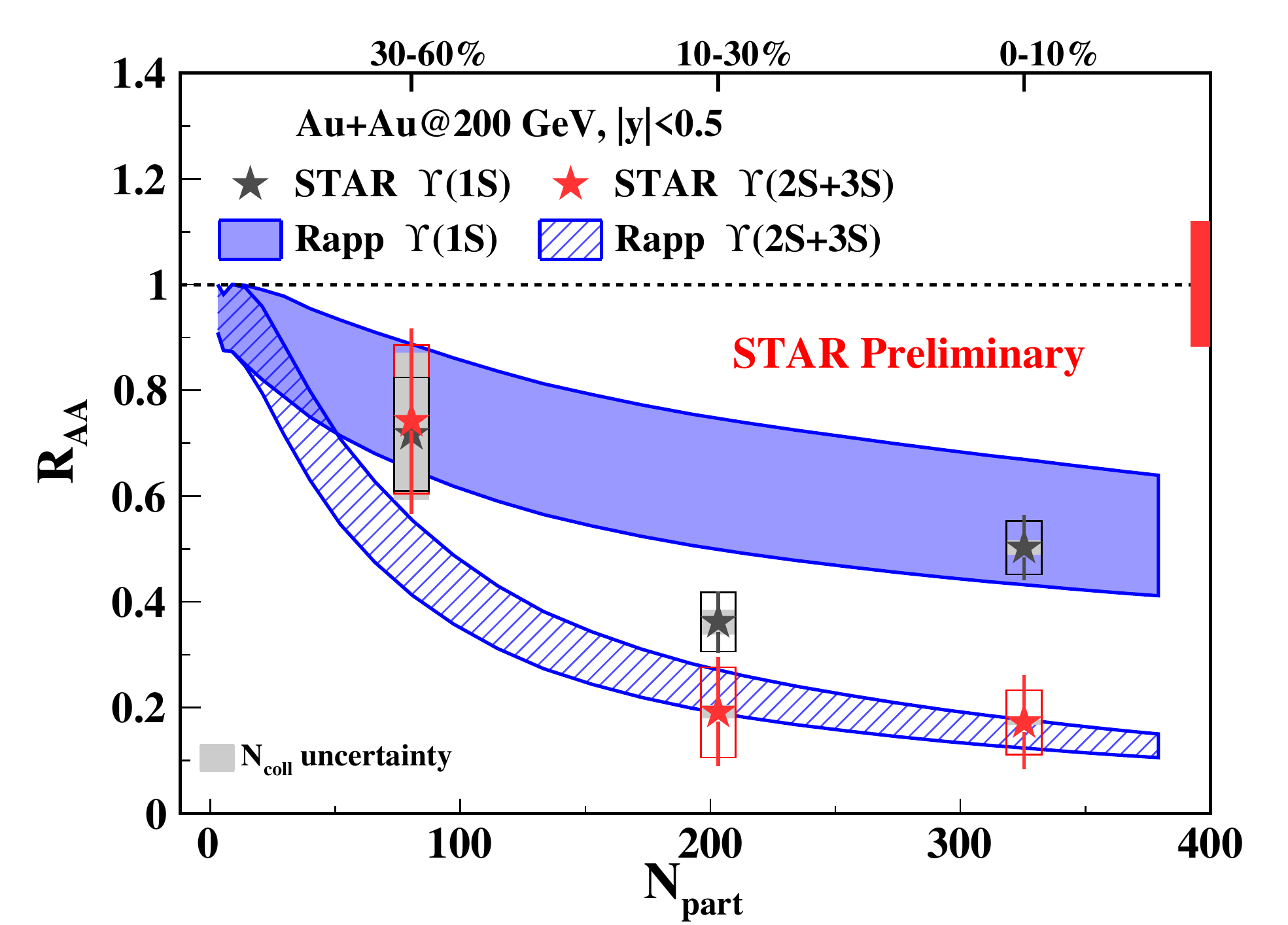}}
\end{minipage}
\begin{minipage}{0.33\linewidth}
\centerline{\includegraphics[width=0.95\linewidth]{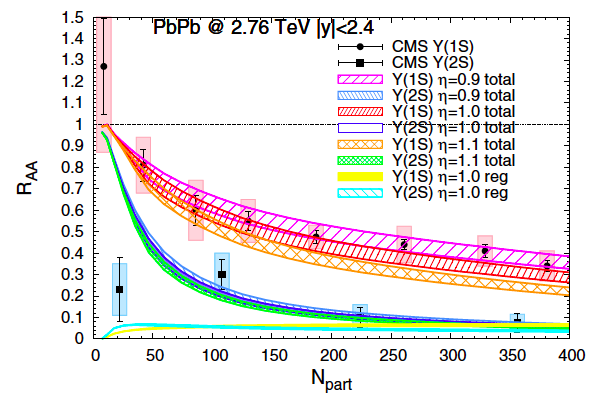}}
\end{minipage}
\begin{minipage}{0.33\linewidth}
\centerline{\includegraphics[width=0.85\linewidth]{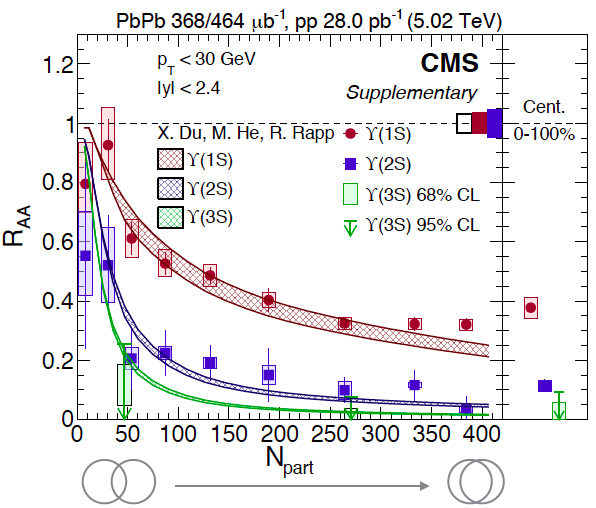}}
\end{minipage}
\caption[]{Measurements of \RAA\ vs. \npart\ for ground and excited \ups\ mesons in \sqrtsNN\ = 200 GeV \AuAu\ (left), \sqrtsNN\ = 2.76 TeV (middle) and 5.02 TeV (right) \PbPb\ collisions. Calculations from a transport model are shown for comparison in all cases \cite{Du:2017qkv}. }
\label{fig:UpsRaa1}
\end{figure}

The three panels of Fig. \ref{fig:UpsRaa1} show the measurements of \ups\ \RAA\ as a function of \npart\ in \sqrtsNN\ = 200 GeV Au+Au collisions from the STAR experiment (left), \sqrtsNN\ = 2.76 TeV (middle) and 5.02 TeV (right) Pb+Pb collisions from the CMS experiment \cite{Khachatryan:2016xxp, Sirunyan:2018nsz}, respectively. In each panel, \RAA\ for the ground and excited states are shown for comparison, where a clear ordering of the suppression level for the different states is seen in central heavy-ion collisions. Compared to the ground \ups\ meson, the excited states of smaller binding energies are more likely to be dissociated in the medium, leading to a stronger suppression. The level of suppression for the inclusive \ups(1S) state is  compatible between \sqrtsNN\ = 200 GeV Au+Au and \sqrtsNN\ = 2.76 TeV Pb+Pb collisions. This could be due to that the inclusive \ups(1S) suppression results mainly from the CNM effects as well as the strong suppression of the excited \ups\ states feeding down to the ground state at these collision energies. The directly generated \ups(1S) remains largely unaltered due to its large binding energy. Furthermore, the inclusive \ups(1S) is slightly more suppressed at \sqrtsNN\ = 5.02 TeV compared to \sqrtsNN\ = 2.76 TeV, indicating maybe an onset of the direct \ups(1S) suppression as the medium temperature reaches higher values at \sqrtsNN\ = 5.02 TeV. On the other hand, the excited \ups\ states seem to be less suppressed at RHIC than those at LHC especially in peripheral collisions. This could be due to that the temperature in peripheral events at RHIC is not even high enough to dissociate the excited states. Comparing the \ups(2S) suppression at \sqrtsNN\ = 2.76 TeV and \sqrtsNN\ = 5.02 TeV, the magnitude is almost the same. One possible explanation is that the medium temperatures at LHC energies are so high that all the \ups(2S) are melted, and only those generated at the phase boundary can be detected experimentally.
%-----------------------------
%-----------------------------
%-----------------------------
\begin{figure}[htbp]
\begin{minipage}{0.33\linewidth}
\centerline{\includegraphics[width=0.95\linewidth]{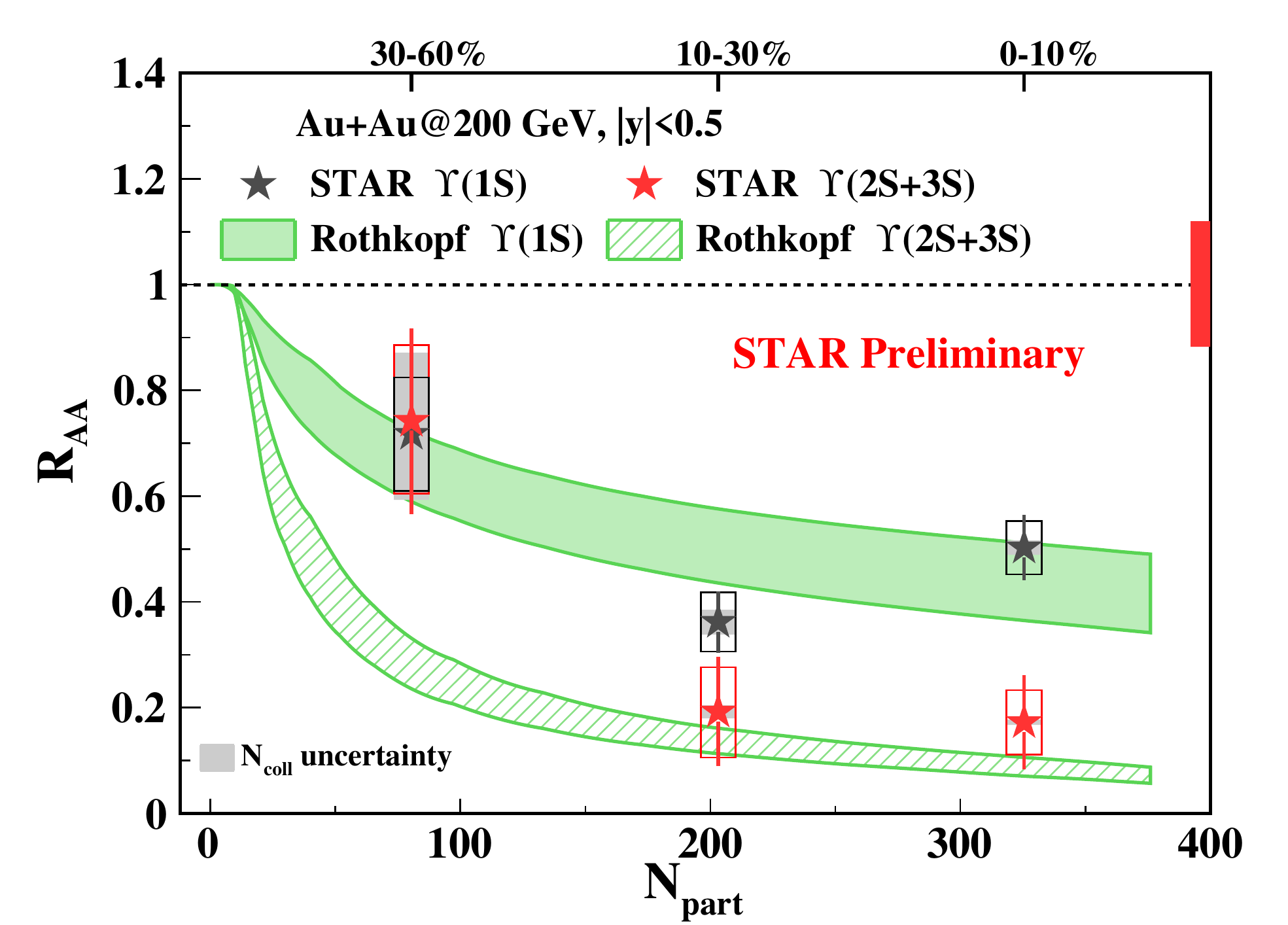}}
\end{minipage}
\begin{minipage}{0.33\linewidth}
\centerline{\includegraphics[width=0.95\linewidth]{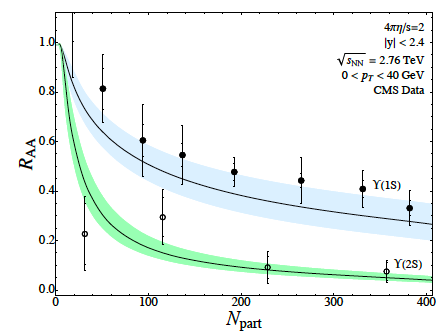}}
\end{minipage}
\begin{minipage}{0.33\linewidth}
\centerline{\includegraphics[width=0.85\linewidth]{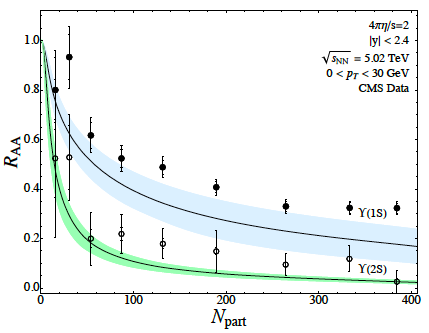}}
\end{minipage}
\caption[]{Measurements of \RAA\ vs. \npart\ as shown in Fig. \ref{fig:UpsRaa1}, but compared to a different model \cite{Krouppa:2017jlg}.}
\label{fig:UpsRaa2}
\end{figure}
%-----------------------------

To further constrain the medium temperature, two model calculations by Du, Rapp, He \cite{Du:2017qkv} and by Krouppa, Rothkopf, Strickland \cite{Krouppa:2017jlg} are compared to experimental data. In the transport model developed by Rapp group, the dissociation and regeneration of the \ups\ mesons are  controlled by a kinetic-rate equation. The temperature-dependent binding energies in the medium are extracted from microscopic T-matrix calculations, while the space-time evolution of the fireball is dictated by a lQCD-based equation of state. The CNM effects are also included. On the other hand, the model by Rothkopf and his collaborators uses a lQCD-based complex value heavy-quark potential coupled with a QGP background following anisotropic hydrodynamic evolution. No regeneration or CNM effects are included. Both model calculations at three different collision energies are shown in Figs. \ref{fig:UpsRaa1} and \ref{fig:UpsRaa2}, respectively. The Rapp model does a reasonably good job in describing the \ups\ \RAA\ across different energies for different states. A small tension is seen for \ups(1S) in very central Pb+Pb collisions at \sqrtsNN\ = 5.02 TeV. The Rothkopf model describes the \ups\ suppression at RHIC fairly well, but seems to consistently lie below the data points from the LHC. In both model calculations, the medium temperatures in central collisions at \sqrtsNN\ = 5.02 TeV exceed the dissociation temperature of the \ups(1S).

\section{Summary}
In these proceedings, recent quarkonium measurements in p+p, p+A and A+A collisions at both RHIC and LHC are presented. Despite the progresses made in understanding the \ups\ production at low \pT, new measurements of \jpsi\ fragmentation function continue to challenge the current theoretical framework. More efforts are needed both theoretically and experimentally to solve the puzzle of the quarkonium production in elementary collisions. In p+A collisions, significant suppression for all quarkonium states is seen at low \pT\ at both mid- and forward rapidities. This needs to be taken into account when interpreting the quarkonium suppression results in heavy-ion collisions. The observation of non-zero \jpsi\ \vtwo\ in intermediate \pT\ region in these collisions calls for further investigation on the origin of the flow sign in light of little yield suppression. In A+A collisions, the strong suppression of high-\pT\ \jpsi\ due to the color screening effect constitutes a strong evidence of the deconfinement. Furthermore, a clear ordering of the suppression pattern for different \ups\ states with different binding energies emerges at both RHIC and LHC, consistent with the sequential suppression scenario. Model calculations including different dissociation temperatures for different \ups\ states can qualitatively describe the experimental data. With the improvement of theoretical calculations and high-precision measurements of the suppression patterns for different quarkonium states at both RHIC and LHC in the future, one aims at constraining and extracting the temperature profile of the medium, and contributing to a full understanding of the QGP properties.



%% References


\bibliographystyle{elsarticle-num}
\bibliography{RMa_qm2018}






% \end{thebibliography}

\end{document}